\documentclass{article}
\usepackage{epsfig}
\begin{document}
\title{Nuclear Stopping as A Probe to In-Medium Nucleon-Nucleon
Cross Section in Intermediate Energy Heavy Ion Collisions }
\author{{\small {Jian-Ye Liu$^{1,2,3}$, Wen-Jun Guo$^{2}$,
Shun-Jin Wang$^{1,3,4}$}}\\ {\small {Wei Zuo$^{1,2,3}$,
Qiang Zhao$^{2}$, Yan-Fang Yang$^{2}$ }} \\
$^{1}${\small Center of Theoretical Nuclear Physics,
  National Laboratory of Heavy}\\
  {\small Ion Accelerator, Lanzhou 730000, P. R. China}\\
$^{2}${\small Institute of Modern Physics, Chinese Academy of
Sciences, Lanzhou 730000, P. R. China}\\
$^{3}${\small CCAST (World Lab.), P.O.Box 8730
  Beijing 100080}\\
$^{4}${\small Department of Modern Physics, Lanzhou
  University, Lanzhou 730000, P. R. China}}
\date{}
\maketitle
\begin{center}
\begin{minipage}{120mm}
\baselineskip 0.3in \small \hskip 0.2in
Using an isospin-dependent quantum molecular dynamics, nuclear stopping
in intermediate heavy ion collisions has been studied.
The calculation has been done for colliding systems with different
neutron-proton ratios in beam energy ranging from 15MeV/u
to 150MeV/u. It is found that, in the energy region from above
Fermi energy to 150MeV/u, nuclear stopping is very sensitive to
the isospin dependence of in-medium nucleon-nucleon cross section,
but insensitive to symmetry potential.
From this investigation, we propose that nuclear stopping can be
used as a new probe to extract the information on the isospin
dependence of in-medium nucleon-nucleon cross section in
intermediate energy heavy ion collisions.
\vskip 0.5in
{\bf PACS number(s):} \hskip 0.05in 25.70.Pq, 25.70.-z, 24.10.Lx
\end{minipage}
\end{center}
\newpage
\baselineskip 0.3in Nuclear stopping in heavy ion collisions (HIC)
has been studied by means of rapidity distribution$^{[1]}$ and
asymmetry of  nucleon momentum distribution$^{[2,3,4]}$. It is an
important quantity in determining the outcome of a
reaction$^{[5,6]}$. W. Bauer et al. pointed out that in
intermediate energy HIC, nuclear stopping power is determined by
both the mean field and the in-medium nucleon-nucleon (N-N) cross
section$^{[4,7]}$, but the mean field he used did not include
symmetry potential. Recently S. A. Bass, S. J. Yennello, H.
Johnston, and Bao-An Li et al suggested that the degree of
approaching to isospin equilibration provides a means to probe
the mechanism and the power of nuclear stopping in HIC
$^{[8-17]}$. But it is not clear how the stopping power depends
on the symmetry potential in the same collisions. In this Letter,
we report a new possibility to extract information on the
in-medium N-N cross section in intermediate energy HIC by using
nuclear stopping as a probe. The effects of both in-medium N-N
cross section and symmetry potential on nuclear stopping shall be
studied comparatively for colliding systems with different
neutron-proton ratios in the beam energy ranging from 15MeV/u to
150MeV/u by using an isospin dependent quantum molecular dynamics
(IQMD) model. The calculated results show that stopping power
depends strongly on the isospin dependence of the in-medium N-N
cross section, but weakly on the symmetry potential for all the
colliding systems studied in the beam energy region from 45MeV/u
to 150MeV/u. However, in the energy region below Fermi energy,
nuclear stopping is sensitive to both the in-medium N-N cross
section and the symmetry potential.
\par
The following two quantities can be used to describe nuclear
stopping in HIC. The momentum quadrupole $Q_{ZZ}$ defined as
$Q_{ZZ}=\sum_{i}^{A}(2P_{z}(i)^{2}-P_{x}(i)^{2}-P_{y}(i)^{2})$,
and the transverse-parallel ratio of momentum $R$ given by $R=
(2/\pi) (\sum_{i}^{A}\mid P_{\perp}(i)\mid)/(\sum_{i}^{A}\mid
P_{\parallel }(i)\mid)$. Here the total mass A is the sum of the
projectile mass $A_p$ and the target mass $A_t$. The values of
the transverse and parallel components of the momentum of the
i-th nucleon are $P_{\perp }(i) =
\sqrt{P_{x}(i)^{2}+P_{y}(i)^{2}}$ and $P_{//}(i)=P_{z}(i)$,
respectively.
\par
In order to describe the isospin effects on the dynamical process
of HIC, quantum molecular dynamics (QMD$^{[11]}$) should be
modified properly: (1) the density dependent mean field should
contain the correct isospin-dependent terms including symmetry
potential and Coulomb potential, (2) the in-medium N-N cross
section should be different for neutron-neutron ( proton-proton )
and neutron-proton collisions, and finally, (3) Pauli blocking
should be counted by distinguishing neutron and proton. The
interaction potential is given by
\begin{equation}
U=U^{Sky}+U^{Yuk}+U^{Coul}+U^{MDI}+U^{Pauli}+U^{Sym},
\end{equation}
where $U^{Sky}$ is the density-dependent Skyrme potential,
\begin{equation}
U^{Sky}=\alpha (\frac \rho {\rho _0})+\beta (\frac \rho {\rho
_0})^\gamma
\end{equation}
$\alpha=-390.1MeV$, $\beta=320.3MeV$ and $\gamma=1.1667$.
$U^{Coul}$ the Coulomb potential and $U^{Yuk}$ the Yukawa
potential (for detail, see Refs.[18]). $U^{MDI}$ is the
momentum-dependent interaction$^{[11]}$,
\begin{equation}
U^{MDI}=t_4ln^2[t_5(\overrightarrow{p_1}-\overrightarrow{p_2})^2+1]\frac{\rho}
{\rho_{0}}
\end{equation}
where the parameters $t_{4}=1.57MeV$,
$t_{5}=5\times10^{-4}MeV^{-2}$, $\rho$ and $\rho_{0}$ are nuclear
density and its normal value, respectively. $U^{Pauli}$ is the
Pauli potential,
\begin{equation}
U^{Pauli}=V_p(\frac \hbar {p_0q_0})^3
exp(-\frac{(\overrightarrow{r_{i}}-
\overrightarrow{r_{j}})^2}{2q_{0}^2}-\frac{(\overrightarrow{p_{i}}-
\overrightarrow{p_{j}})^2}{2p_{0}^2})\delta _{p_{i}p_{j}}
\end{equation}
where $\delta _{p_ip_j}=1$ for neutron-neutron or proton-proton,
and $\delta _{p_ip_j}=0$ for neutron-proton. The parameters
$V_{p}=30MeV$, $p_{0}=400MeV/c$ and $q_{0}=5.64fm$.
 According to our experience, the interaction potentials containing
Pauli potential can describe the structure effect of fragmentation
in the dynamical process of HIC $^{[19]}$.
\par
$U^{sym}$ is the symmetry potential. In the present calculation,
three different density dependences of the symmetry potential
$^{[12]}$ are used, i.e., $U_1^{sym}=cF_1(u)\delta \tau _z$,
$U_2^{sym}=cF_2(u)[\delta \tau _z+\frac{1}{2}\delta ^2]$, and
$U_3^{sym}=cF_3(u)[\delta \tau _z-\frac {1}{4}\delta ^2]$, where
$\tau_{z}=1$ for neutron and $\tau_{z}=-1$ for proton, c is the
strength of symmetry potential, taking the value of 0 or 32MeV.
$F_1(u)=u$, $F_2(u)=u^2$ and $F_3(u)=u^{1/2}$, $u\equiv \rho/
\rho_{0}$. $\delta$ is the relative neutron excess $\delta
=\frac{\rho_n-\rho_p}{\rho_n+\rho_p}
        =\frac{\rho_n-\rho_p}{\rho} $,
where $\rho_n$ and $\rho_p$ are neutron and proton densities,
respectively. First of all, the density distribution of nucleus
was calculated by using Skyrme-Hatree-Fock with parameter set
SKM*$^{[22]}$.
\par
An empirical expression of the in-medium N-N cross section
$^{[20]}$ is used, $\sigma _{NN}^{med}=(1+\alpha \frac \rho {\rho
_0})\sigma _{NN}^{free}$, with the parameter $\alpha \approx
-0.2$, and $\sigma_{NN}^{free}$ is the experimental free N-N
cross section$^{[21]}$. It is known that the free neutron-proton
cross section is about 3 times larger than the free proton-proton
or neutron-neutron cross section below 300 MeV.
\par
The nuclear stopping are studied for the colliding systems
$^{20}Ne + ^{20}Ne$, $^{40}Ar + ^{40}Ar$, $^{80}Zn + ^{80}Zn$,
$^{112}Sn + ^{112}Sn$ and $^{124}Sn + ^{124}Sn$ at different beam
energies ranging from 15MeV/u to 150 MeV/u. The neutron-proton
ratios of the above colliding systems are 1.0, 1.22, 1.67, 1.24,
and 1.48, respectively. To make a comparative study of the
isospin effects of the in-medium N-N cross section and the
symmetry potentials, we have investigated the following three
different cases. (i) The symmetry potential $U^{sym} = U_1^{sym}$
is used with the strength $c=32$ MeV and the isospin dependent
in-medium N-N cross section is employed. This case is denoted by
C=$U_1^{sym}$ + $\sigma^{iso}$ and solid lines in the figures.
(ii) $ U^{sym} = U_1^{sym}$ and $\sigma_{NN}^{med}$ is
isospin-independent, denoted by C=$U_1^{sym}$ + $\sigma^{noiso}$
and dash lines. (iii) There is no symmetry potential ($c=0$) and
$\sigma_{NN}^{med}$ is isospin-dependent, denoted by C=0 +
$\sigma^{iso}$ and dotted lines.
\par
Fig.1 depicts the time evolution of $R$ for central collisions of
the systems $^{112}Sn + ^{112}Sn$ (top row), and $^{124}Sn +
^{124}Sn$ (bottom row), with the beam energies of 15MeV/u,
25MeV/u, 35MeV/u, 45MeV/u, 72MeV/u, 100MeV/u, and 150MeV/u for the
above three cases. It is very clear that for both colliding
systems the nuclear stopping $R$ depends strongly on the isospin
dependence of the in-medium N-N cross section and weakly on the
symmetry potential as the beam energy is above 45MeV/u. As the
beam energy decreases to below the Fermi energy, the nuclear
stopping depends on both isospin dependence of in-medium N-N
cross section and the symmetry potential.
\par
In Fig.~2, the time evolution of $Q_{ZZ}$ is shown for central
collisions of the system $^{112}$Sn + $^{112}$Sn at different beam
energies in the case of $C=U_1^{sym}$ + $\sigma^{iso}$. It is
clearly shown that as decreasing the beam energy, the asymptotic
value of $Q_{ZZ}$ decreases towards zero, indicating an isotropic
nucleon momentum distribution of the whole composite system and
consequently a full stopping at the beam energy below Fermi
energy. As beam energy increases to above Fermi energy, because
of the pre-equilibrium particle emission and the longitudinal
motion of the project-like and target-like nuclei, $Q_{zz}$ will
get certain non-vanishing value, indicating partial transparency.
It is also seen that the relaxation time decreases as increasing
beam energy, which shows the fact that high beam energy leads to
more violent N-N collisions and faster dissipation. This is
consistent with the isospin equilibrium process as shown by Bao-An
Li et al. $^{[12]}$.

\par
In order to trace neutron and/or proton observables, in Fig.~3 is
given the time evolution of $Q_{zz}$ for neutron and proton
respectively. The conclusion drawn from Fig.~1 is applicable to
both neutron and proton stoppings.
\par
In Fig.~4 is plotted the time evolution of $R$ for the reaction
$^{112}$Sn + $^{112}$Sn at $E=100$MeV/u for different impact
parameters. It is noticed that the nuclear stopping for small
impact parameters (b=0.0fm, 1.0fm, 2.0fm) shows the same behavior
as the central collisions as in Fig.~1, while as increasing the
impact parameter, the behavior becomes different and the dominant
role played by the isospin dependent in-medium N-N cross section
gradually disappears. The impact parameter dependence of $R(b)$
can be seen more clearly from Table 1 where the asymptotic values
of $R$ are given for different $b$.

\begin{center}
\begin{minipage}{120mm} {\bf Table 1} The asymptotic values of the
stopping R for the reaction $^{112}Sn+^{112}Sn$ at E=100MeV/u as
function of impact parameter
\end{minipage}
\begin{tabular}{|c|c|c|c|c|c|c|}
\hline
  b(fm) & 0.0 & 1.0 & 2.0 & 3.0 & 4.0 & 5.0 \\ \hline
  $R(U^{sym}_{1}+\sigma^{noiso})$ & 0.70 & 0.72 & 0.60 & 0.55 &
  0.51 & 0.50 \\ \hline
  $R(U^{sym}_{1}+\sigma^{iso})$ & 0.82 & 0.82 & 0.71 & 0.63 &
  0.56 & 0.50 \\ \hline
  $R(0+\sigma^{iso})$ & 0.84 & 0.81 & 0.74 & 0.64 &
  0.58 & 0.51 \\ \hline
\end{tabular}
\end{center}
\par
In Fig.5 is shown the impact parameter-averaged asymptotic values
( the values at $ t \ge 200 $fm/c when the nuclear stopping
becomes nearly a constant as shown in Figs.~1 and 2) of $Q_{zz}$
per nucleon (left window) and $R$ (right window) as a function of
$A_{T}+A_{P}$ in the following seven cases: C=0 + $\sigma^{iso}$
by the solid square; C=$U_1^{sym}$ + $\sigma^{iso}$ denoted by
the solid circles; C=$U_2^{sym}$ + $\sigma^{iso}$ by solid
triangle; C=$U_3^{sym}$ + $\sigma^{iso}$ by solid diamond;
C=$U_1^{sym}$ + $\sigma^{noiso}$ by open circles; C=$U_2^{sym}$ +
$\sigma^{noiso}$ by open triangle; C=$U_3^{sym}$ +
$\sigma^{noiso}$ by open diamond. In the calculation the beam
energy is 100 MeV/u and five colliding systems have been
considered, i.e., $^{20}Ne + ^{20}Ne$, $^{40}Ar + ^{40}Ar$,
$^{80}Zn + ^{80}Zn$, $^{112}Sn + ^{112}Sn$ and $^{124}Sn +
^{124}Sn$. From Fig.~5, it can be seen that the nuclear stopping
depends strongly on the isospin dependence of in-medium N-N cross
section and weakly on the symmetry potential, though it is
slightly different for the three different forms of symmetry
potential, especially for larger-mass systems. It is also shown
that the nuclear stopping power increases as increasing the total
mass of the colliding system, the decrease of $Q_{zz}$ or the
increase of $R$ implies the increase of nuclear stopping.
\par
Fig.~6 shows the asymptotic value at $t=200$fm/c of $R$ as a
function of the beam energy for the same colliding systems and
the same collision conditions as in Fig.~1. From the figure, one
can draw the same conclusion as from Fig.~1.
\par
From the above comparative study of nuclear stopping for different
colliding systems, beam energies, and impact parameters by using
IQMD, we have obtained the following physical picture of the
collision dynamics: from above the Fermi energy to about
150MeV/u, the dynamics is dominated by N-N collisions and the
role played by the mean field is less important. The main
consequence of the N-N collisions is the transformation of the
initial longitudinal motion to the motion in other directions and
the subsequent thermalization of the system. In this case,
nuclear stopping $R$ and $Q_{zz}$ are sensitive to
nucleon-nucleon collisions, and can thus be used as a measure of
the dissipative process. However, as the beam energy decreases to
below the Fermi energy, the collision dynamics is governed by
both the mean field and the N-N collision which result in
one-body and two-body dissipation or thermalization,
respectively. Thus as indicators of the thermalization, $R$ and
$Q_{zz}$ depend on both the mean field and the in-medium N-N
cross-section.
\par
In summary, from the above results and discussions we conclude that
the nuclear stopping $R$ and $Q_{zz}$ can be used as a probe to extract
information on the isospin dependence of the in-medium N-N cross section
in HIC in the beam energy region from above the Fermi energy to about
150 MeV/u.

\section*{Acknowledgment}
\hskip 0.3in This work was supported in part by the `` 100 person
project '' of the Chinese Academy of Sciences, National 973
subject under grant No. G2000077400, the National Natural
Foundation of China under Grants No. 19775057 and No. 19775020,
No. 19847002, No. 19775052 and KJ951-A1-410, by the Foundation of
the Chinese Academy of Sciences.

\newpage

\begin{figure}[htb]
\begin{center}
\hspace*{6cm}
\epsfig{figure=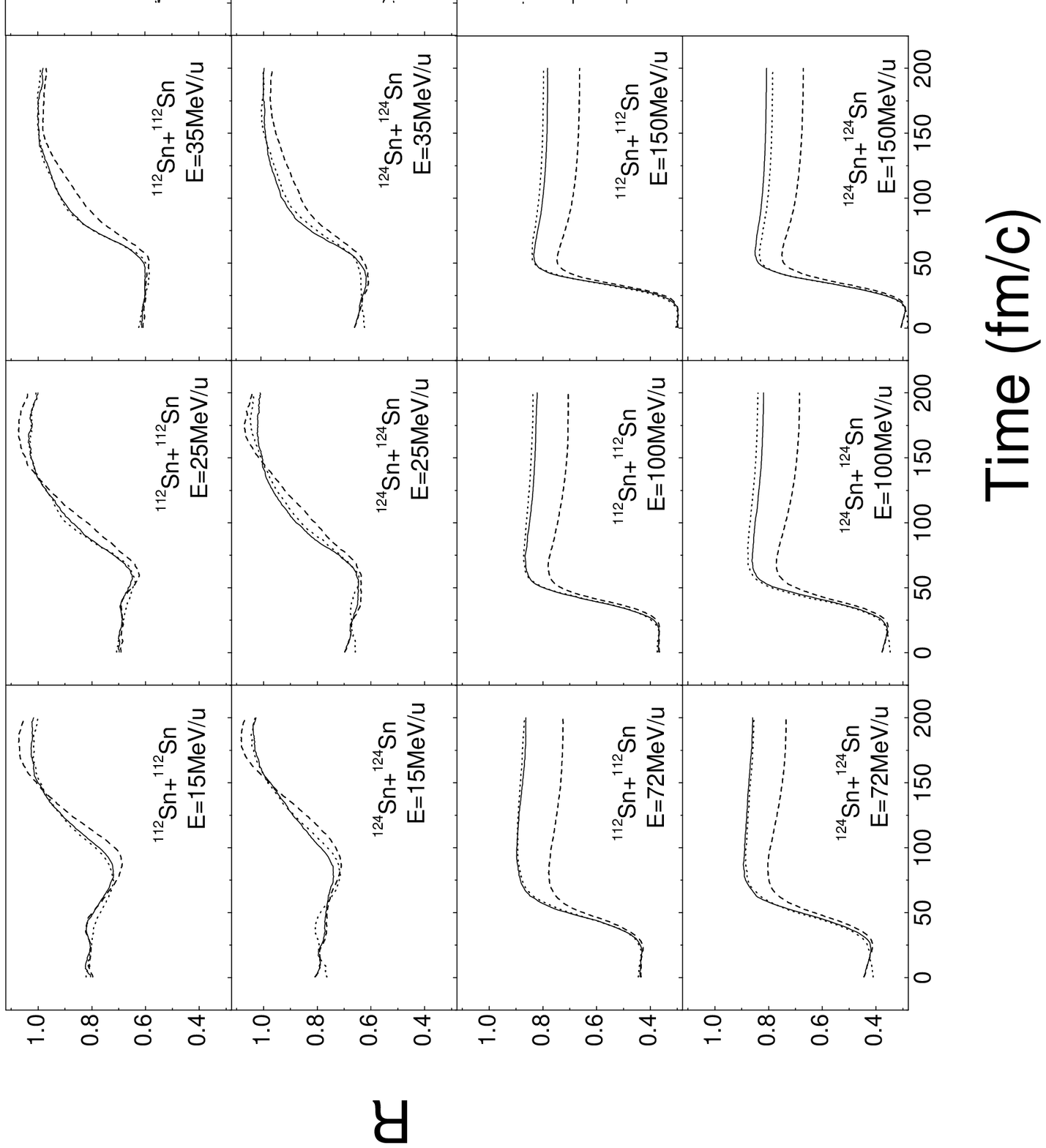, angle=90, height=20cm, width =17cm}
\end{center}
\caption[]{The time evolution of $R$ in central collisions for
two colliding systems  $^{112}$Sn +$^{112}$Sn, and $^{124}$Sn +
$^{124}$Sn in the three cases (see the text). }
\end{figure}

\newpage

\begin{figure}[htb]
\begin{center}
\hspace*{6cm}
\epsfig{figure=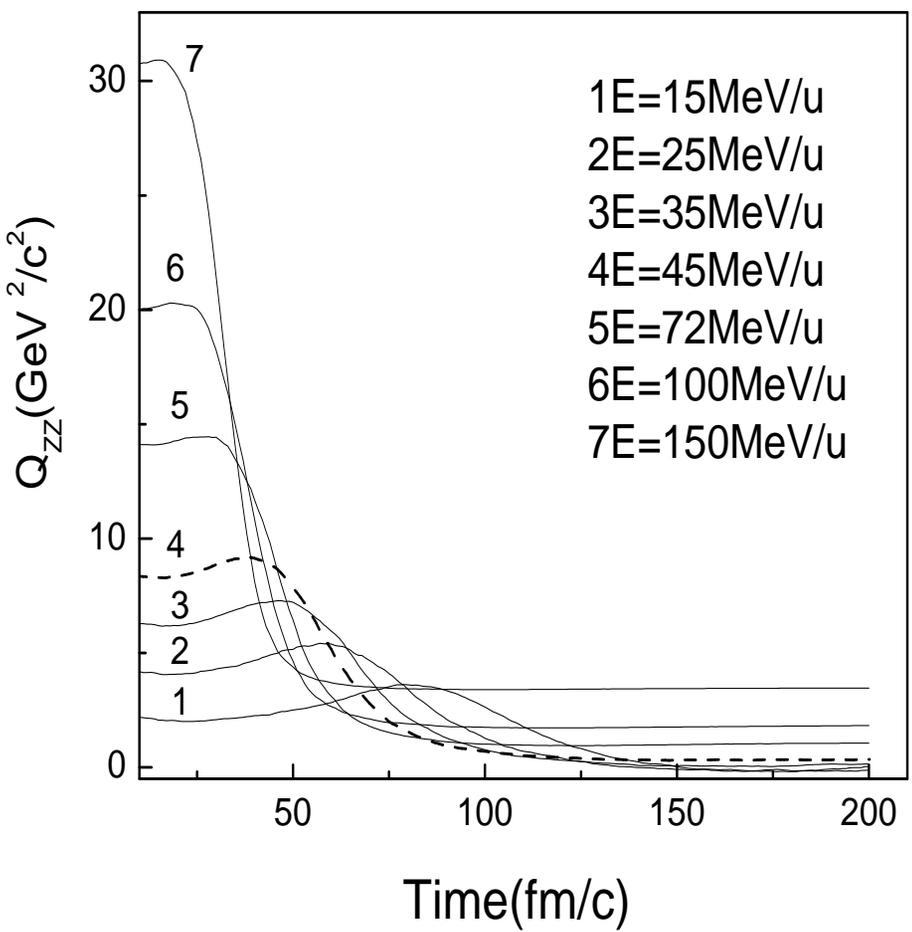, angle=90, height=20cm, width =16cm}
\end{center}
\caption[]{ The time evolution of $Q_{ZZ}$ in central collisions
of  $^{112}$Sn +$^{112}$Sn  for different beam energies in the
case of C=$U_1^{sym}$+$\sigma^{iso}$. }
\end{figure}

\newpage

\begin{figure}[htb]
\begin{center}
\hspace*{6cm}
\epsfig{figure=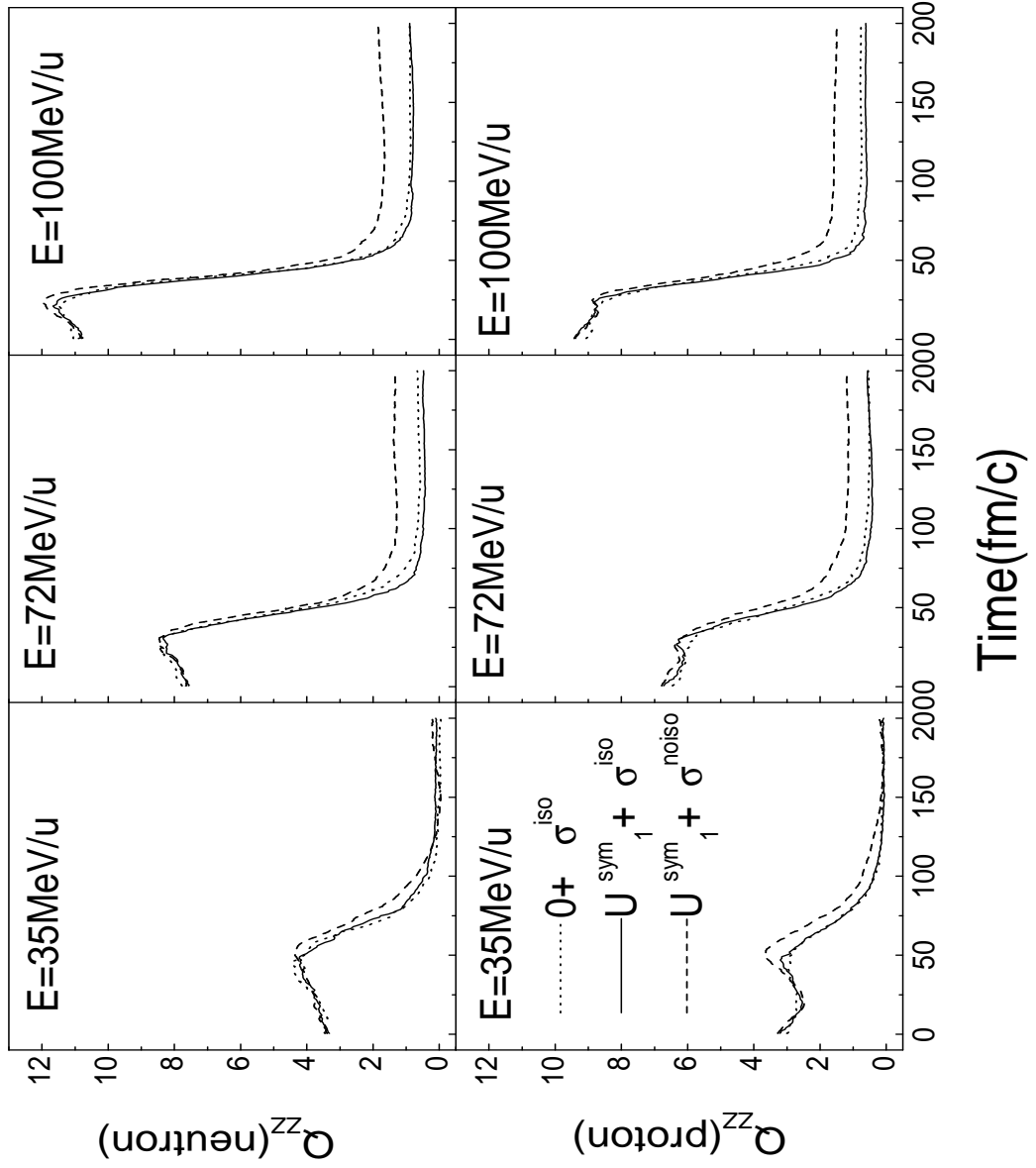, angle=90, height=20cm, width =16cm}
\end{center}
\caption[]{ The time evolution of momentum quadrupoles for neutron
$Q_{zz}(neutron)$ and proton $Q_{zz}(proton)$ for the central
collisions of $^{112}Sn + ^{112}Sn$ at the beam energies of
35MeV/u, 72MeV/u, and 100MeV/u in the three cases as shown in
Fig.1. }
\end{figure}

\newpage

\begin{figure}[htb]
\begin{center}
\hspace*{6cm}
\epsfig{figure=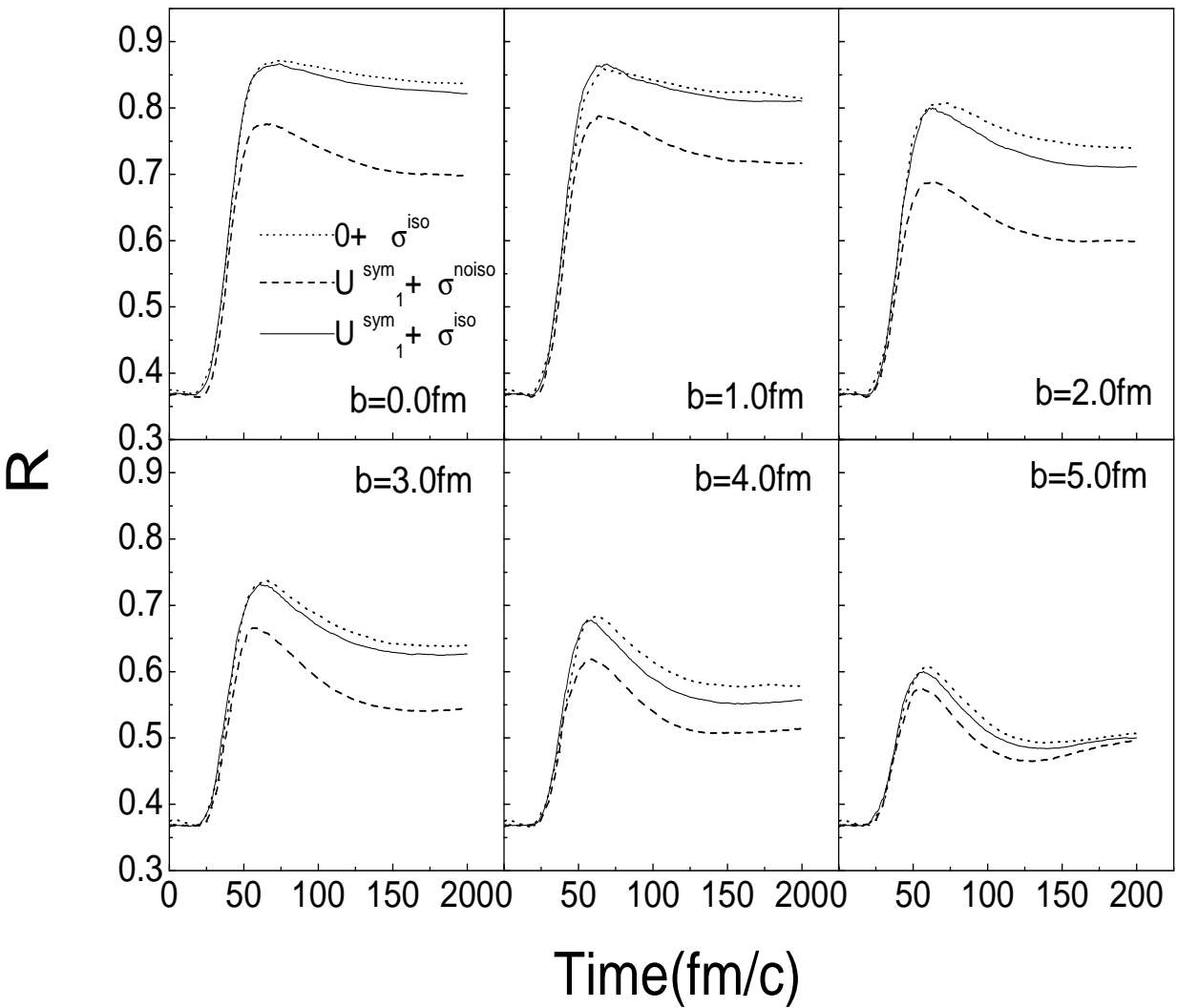, angle=90, height=20cm, width =16cm}
\end{center}
\caption[]{ The time evolution of  $R$ at different
impact parameters for system $^{112}Sn + ^{112}Sn$ at E=100MeV/u
in the three cases as shown in Fig.1. }
\end{figure}

\newpage

\begin{figure}[htb]
\begin{center}
\hspace*{6cm}
\epsfig{figure=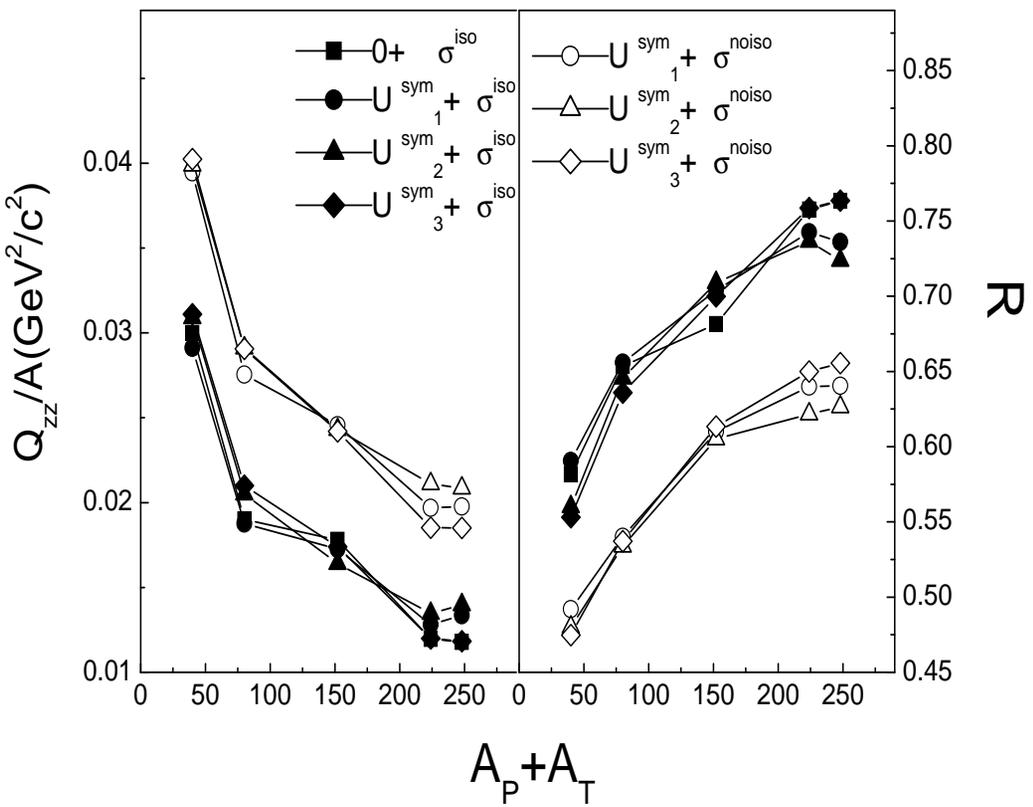, angle=90, height=20cm, width=16cm}
\end{center}
\caption[]{ The impact parameter averaged asymptotic value of
$Q_{zz}$ (left window) and $R$ (right window) as a function of
$A_{T}+A_{P}$ at E=100MeV/u in the seven cases (see the text). }
\end{figure}

\newpage

\begin{figure}[htb]
\begin{center}
\hspace*{6cm}
\epsfig{figure=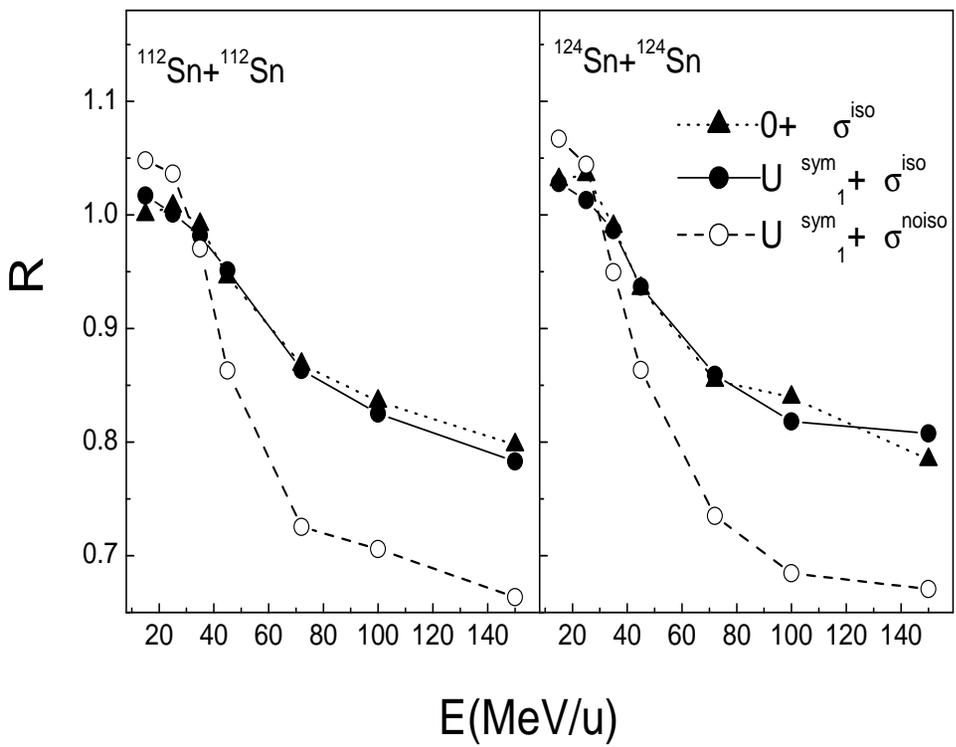, angle=90, height=20cm, width=16cm}
\end{center}
\caption[]{ The nuclear stopping $R$ as a function of beam
energy for two collision systems at t=200fm/c in the three cases
as in Fig.1. } 
\end{figure}

\end{document}